\documentclass[reprint,aip,superscriptaddress,longbibliography]{revtex4-1}

\usepackage{graphicx}
\usepackage{amsmath}
\usepackage{amsfonts}
\usepackage{amssymb}

\newcommand{\DZPmv}{$^{(P)}\mathrm{d}\zeta+\mathrm{p}$}

\begin{document}

\title{Structural and configurational properties of nanoconfined monolayer ice from first principles}
\author{Fabiano~Corsetti}
\email[E-mail: ]{fabiano.corsetti08@imperial.ac.uk}
\affiliation{CIC nanoGUNE, 20018 Donostia-San Sebasti\'{a}n, Spain}
\affiliation{Department of Materials and the Thomas Young Centre for Theory and Simulation of Materials, Imperial College London, London SW7 2AZ, United Kingdom}
\author{Paul~Matthews}
\affiliation{CIC nanoGUNE, 20018 Donostia-San Sebasti\'{a}n, Spain}
\author{Emilio~Artacho}
\affiliation{CIC nanoGUNE, 20018 Donostia-San Sebasti\'{a}n, Spain}
\affiliation{Theory of Condensed Matter, Cavendish Laboratory, University of Cambridge, Cambridge CB3 0HE, United Kingdom}
\affiliation{Basque Foundation for Science Ikerbasque, 48011 Bilbao, Spain}
\affiliation{Donostia International Physics Center, 20018 Donostia-San Sebasti\'{a}n, Spain}

\begin{abstract}
Understanding the structural tendencies of nanoconfined water is of great interest for nanoscience and biology, where nano/micro-sized objects may be separated by very few layers of water. Here we investigate the properties of ice confined to a quasi-2D monolayer by a featureless, chemically neutral potential, in order to characterize its intrinsic behaviour. We use density-functional theory simulations with a non-local van der Waals density functional. An {\em ab initio} random structure search reveals all the energetically competitive monolayer configurations to belong to only two of the previously-identified families, characterized by a square or honeycomb hydrogen-bonding network, respectively. We discuss the modified ice rules needed for each network, and propose a simple point dipole 2D lattice model that successfully explains the energetics of the square configurations. All identified stable phases for both networks are found to be non-polar (but with a topologically non-trivial texture for the square) and, hence, non-ferroelectric, in contrast to previous predictions from a five-site empirical force-field model. Our results are in good agreement with very recently reported experimental observations.
\end{abstract}

\maketitle

Thanks to its open and flexible tetrahedral network, water is famous for the large degree of polymorphism it exhibits in the crystal state\cite{Soper23082002, 0036-021X-75-1-R04}, also having interesting ramifications for the amorphous solid and, more controversially, the liquid\cite{Mishima1985, Mishima1998, Brovchenko2008, Huang2009, Kesselring2012, Holten2013, Limmer01072014}. One surprisingly constant feature of all the ice phases (at least until very high pressures\cite{PhysRevLett.81.1235, Cavazzoni01011999, Pickard2013, PhysRevB.84.220105}), is the obedience of the so-called ice rules originally described by Bernal and Fowler\cite{Bernal1933}, by which each water molecule has four hydrogen-bonded neighbours (in a preferentially quasi-tetrahedral configuration) with two short OH distances (the donated protons) and two long ones (the accepted protons). This asymmetry in the hydrogen bond gives rise to the possibility of disordered phases with a residual entropy, the most common form of ice (Ih) being a notable example\cite{Pauling1935, Kolafa2014}.

The behaviour of water at surfaces\cite{Yang2004, Feibelman2004, Carrasco2012, Li2012, Kaya2013, Cheh2013} and nanoconfined in three\cite{Rates2011}, two\cite{Koga2001, Hummer2001, Takaiwa08012008, Kyakuno2010} or one dimensions\cite{Lee1984, Koga1997, Meyer1999, Koga2000, Zangi2003, Zangi2003a, Kumar2005, Koga2005, Giovambattista2006, Kumar2007, Giovambattista2009, Johnston2010, Han2010, Bai2010, Mazza2012, Mosaddeghi2012, Ferguson2012, Qiu2013, Kaneko2013, Zhao2014, Kaneko2014, Zhao2014a, bilayer-new} is of importance for numerous biological, geological, and industrial systems, as well as being a useful testing ground for learning about the intrinsic properties of water itself\cite{Kumar2005, Han2010, Mazza2012}. In particular, several theoretical studies\cite{Zangi2003a, Kumar2005, Koga2005, Bai2010, Mosaddeghi2012, Ferguson2012, Qiu2013, Kaneko2013, Zhao2014, Kaneko2014, Zhao2014a} have focussed on the properties of a quasi-2D monolayer of water confined in one dimension between infinite plates. The confinement width needed for a stable monolayer is very small, no more than $\sim$7--8~\AA\ based on these previous studies (this is of course somewhat dependent on the details of the confining walls).

\begin{figure*}[t!]
\includegraphics[width=0.98\textwidth]{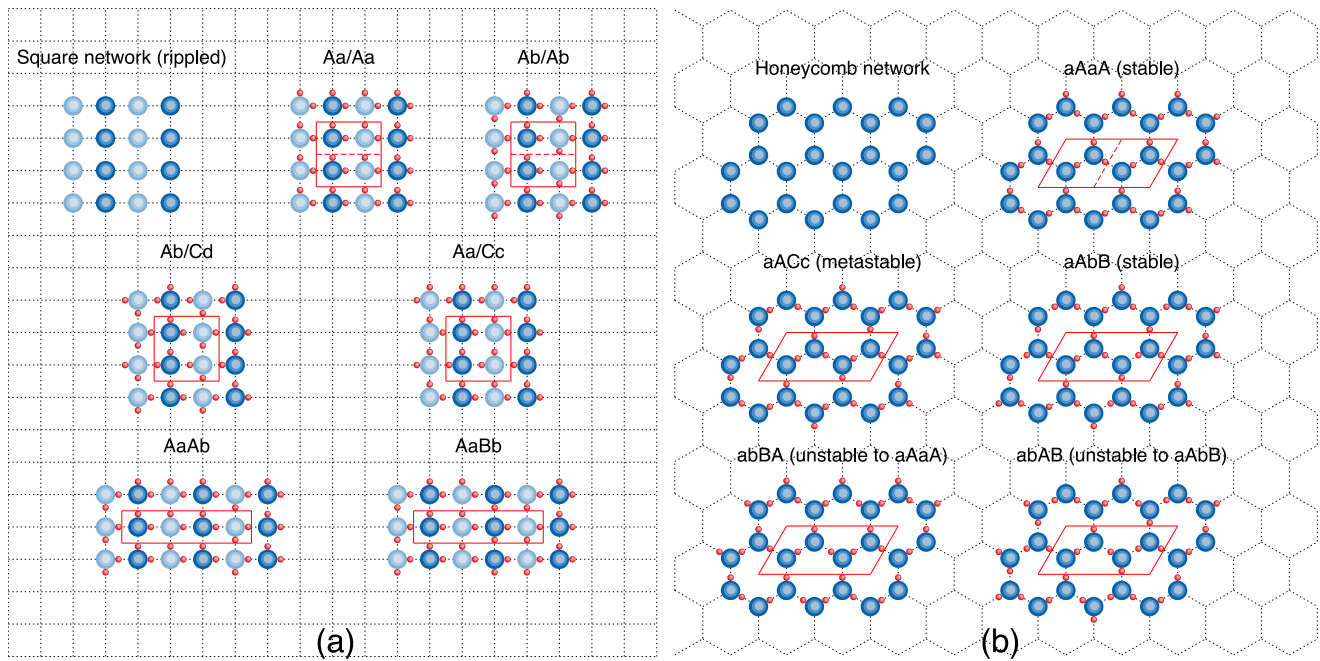}
\caption{{\bf Schematic diagrams for the two different families of configurations.} All allowed non-equivalent configurations for a unit cell of four water molecules are shown for each network. The lattice vectors are shown in red; if a smaller unit cell is possible, this is shown by a dashed red line. For (a), upper/lower case letters indicate O ions in different z positions (also shown by the two different shadings). The letters indicate the four possible orientations of the water molecule on the lattice. For (b), upper/lower case letters indicate whether the water molecule is in/out of plane; in the latter case, only the proton in plane is shown, while the other one can be either above or below the O ion. The letters indicate the three possible orientations of the water molecule in plane.}
\label{fig:configs_diag}
\end{figure*}

The particularly interesting feature of the monolayer is that it is not possible to respect both the bulk ice rules and the preferential tetrahedral bonding geometry. So, how do the molecules arrange? A comprehensive overview of the predictions obtained so far from simulation is given by Zhao {\em et al.}\cite{Zhao2014a} in a very recent paper. A number of different monolayer ice phases have been identified; in general, the four-fold coordination rule appears to be satisfied, although the two lowest density phases, a truncated square tiling (the so-called Archimedean $4 \cdot 8^2$) and an elongated hexagonal, feature a novel dimer configuration of pairs of molecules connected by two hydrogen bonds (which are therefore extremely distorted)\cite{Zhao2014}.

The most ubiquitous family of monolayer ice configurations, reported in nearly all studies\cite{Zangi2003a, Kumar2005, Koga2005, Bai2010, Mosaddeghi2012, Qiu2013, Kaneko2013, Zhao2014, Kaneko2014}, are those characterized by a square bonding network topology, in which the four-fold coordination of each water molecule is always unambiguously satisfied (Fig.~\ref{fig:configs_diag}(a)); the crystal symmetry, however, is mostly found to be rhombic instead of square due to a distortion of the unit cell. These configurations are higher in density than the Archimedean tiling and hexagonal phases, and are therefore expected to be stabilized at high lateral pressure. Furthermore, there appears to be a transition between an almost completely flat rhombic phase, which competes with the low-density phases for small confinement widths, and a rippled rhombic phase, which is instead stable over almost all of the pressure range for larger confinement widths\cite{Kaneko2013, Zhao2014, Kaneko2014, Zhao2014a}.

Another interesting question is the ferroelectricity of monolayer ice. Both the hexagonal and the flat/rippled rhombic phases have been predicted to be ferroelectric by Zhao {\em et al.}\cite{Zhao2014}. A key prerequisite for this is that the phases are {\em ordered}, and that the most stable configuration in each case is one with a net polarization. Such a conclusion is supported for the rippled rhombic phase by the findings of Zangi and Mark\cite{Zangi2003a}; however, other studies report disordered quasi-square\cite{Kumar2005, Koga2005, Bai2010, Mosaddeghi2012, Qiu2013, Kaneko2013, Kaneko2014} or hexagonal\cite{Ferguson2012} phases.

Very recently, experimental observations of monolayer ice have been presented\cite{Algara-Siller2015}, showing a strictly square lattice of O ions. The overall picture from all previous computational studies does not provide clear support for this finding, and a physical explanation of the relative stability of the square and rhombic crystal symmetries is yet lacking.

The simulations undertaken so far have made use of well-established empirical force-field models of water: TIP4P\cite{Koga2005, Kaneko2013, Kaneko2014}, TIP5P\cite{Zangi2003a, Kumar2005, Bai2010, Qiu2013, Zhao2014, Zhao2014a}, and SPC/E\cite{Mosaddeghi2012, Ferguson2012, Algara-Siller2015}. Zhao {\em et al.}\cite{Zhao2014, Zhao2014a} have performed some additional checks with all of these models plus others (SPC, TIP3P, TIP4P/2005), as well as with first principles calculations using density-functional theory (DFT) with a semi-local (PBE-GGA\cite{pbe}) exchange and correlation (xc) functional. However, the DFT calculations were only carried out for a few examples in order to confirm the structural stability of the phases found by TIP5P; furthermore, these were not tested with the same confining potential, but either free-standing or confined between graphene sheets.

Most of the development and parameterization of the various empirical models has focussed on reproducing characteristics of the bulk phases (even for the bulk, however, there are known to be important discrepancies between them\cite{Vega2009}); their accuracy in a very different and extreme environment such as that considered here is an open question. Therefore, this is an area in which the predictive power of first principles calculations can provide valuable information. This is especially true thanks to the development of fully non-local xc functionals that account for van der Waals (vdW) interactions from first principles\cite{Dion2004}; these have helped to overcome the severe problems previously encountered by (semi-)local functionals in describing the properties of water, and have given a series of very promising results both for liquid water\cite{Mogelhoj2011, Zhang2011a, Corsetti2013b} and ice\cite{Pamuk2012, Murray2012}.

In this paper, we investigate the phase diagram of nanoconfined monolayer ice entirely {\em ab initio}, using DFT with a non-local vdW xc functional. We do not base ourselves on previously identified phases, rather employing a preliminary {\em ab initio} random structure search\cite{Pickard2011} (AIRSS) procedure to identify promising low-enthalpy structures. The questions we aim to answer are: What are the intrinsically stable phases of monolayer ice over a range of confinement widths and lateral pressures? Are these phases ordered or disordered? What contribution do configurational entropy and vibrational effects have on phase stability? What factors determine the stability of individual configurations? Finally, is the electronic structure of the water molecule altered by confinement?

\section*{Results}

\noindent {\bf {\em Ab initio} random structure search.} In order to identify configurations of interest in an unbiased way, we begin our investigation by performing an AIRSS on the confined geometry. For a given value of the confinement width $d$ and the 2D lateral pressure $P_l$ (see definitions in the Methods), we generate many trial unit cells, which are then relaxed. We use 100 unit cells of four molecules and 200 unit cells of eight molecules; the length of the two lattice vectors in the $x$--$y$ plane and the angle between them are set randomly (within a sensible range), and then the water molecules are placed inside the confinement region with random positions and orientations. No symmetry is imposed on the system. We do this at each point on a regular grid on the $P_l$--$d$ phase diagram, for $d$ from 5 to 10~\AA\ in steps of 1~\AA, and $P_l$ from 0.01 to 1000~GPa$\cdot$\AA\ logarithmically in order of magnitude steps. For each point, we rank the 300 relaxed configurations by enthalpy/molecule ($H = U + P_l A$).

The first important result is that stable monolayer configurations are only recovered for $d \le 7$~\AA\ and $P_l \le 10$~GPa$\cdot$\AA. Therefore, we focus on this reduced part of the phase diagram, and leave the analysis of multilayer ices for the future\cite{bilayer-us}.

It is noticeable that, for {\em all} grid points in the monolayer region, the lowest-enthalpy structure found by our search is a square configuration (as explained previously, this refers to the network topology based on hydrogen bonding, and not necessarily the symmetry classification of the ionic structure, discussed below). Furthermore, for the points at $d = 6$~\AA\ and $d = 7$~\AA, the majority (75\%) of structures within 50~meV of the lowest one belong to this square network (SN) family; the remaining configurations are SNs with bonding defects (20\%), and a small number of bilayer configurations (5\%) in the 40--50 meV range.

Instead, for the points at $d = 5$~\AA, more variety is found within the 50~meV range: the number of non-SN configurations ranges from $\sim$20\% of the total at $P_l = 10$~GPa$\cdot$\AA\ up to $\sim$90\% at $P_l = 0.01$~GPa$\cdot$\AA. These can be classified into four types: honeycomb network (HN) configurations, amorphous configurations (made up of irregular combinations of squares, pentagons, hexagons, and some larger voids), SNs with bonding defects, and sub-2D networks (1D chains and ribbons, 0D square islands). The latter type can also be understood as SNs with systematic bonding defects, as the O ions are found to form quasi-square lattices. The HN and amorphous configurations are found in the higher energy range (20--50 meV above the SN ground state).\\

\noindent {\bf Configurational energetics and entropy.} The AIRSS seems to indicate quite definitely the square as the most stable bonding network over the entire monolayer range. However, as with the bulk phases of ice, this does not correspond to a single ionic arrangement, even neglecting the possibility of bonding defects. The same is true of the HN. More importantly, such an analysis cannot tell us about the crucial entropic contribution. We therefore turn to a systematic investigation of the possible configurations for the two networks.

\begin{figure}[t!]
\includegraphics[width=0.49\textwidth]{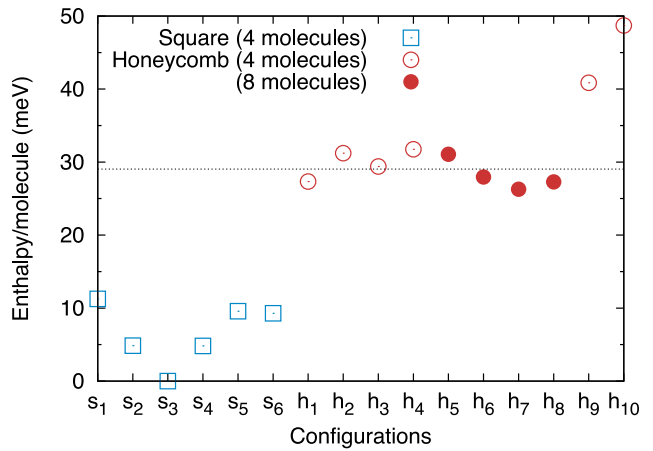}
\caption{{\bf Enthalpy of different configurations for the two networks.} The values shown are for $d = 5$~\AA, $P_l = 0.01$~GPa$\cdot$\AA. The configurations are as follows (using the nomenclature of Fig.~\ref{fig:configs_diag}, and dashes above/below letters to indicate dangling protons above/below the plane): for the SN, s$_1 =$~Aa/Aa, s$_2 =$~Ab/Ab, s$_3 =$~Ab/Cd, s$_4 =$~Aa/Cc, s$_5 =$~AaAb, s$_6 =$~AaBb; for the HN, h$_1 =$~$\overline{\mathrm{a}}$A\underline{a}A, h$_2 =$~$\overline{\mathrm{a}}$A$\overline{\mathrm{a}}$A, h$_3 =$~$\overline{\mathrm{a}}$A\underline{b}B, h$_4 =$~$\overline{\mathrm{a}}$A$\overline{\mathrm{b}}$B, h$_5 =$~$\overline{\mathrm{a}}$A$\overline{\mathrm{a}}$A/$\overline{\mathrm{a}}$A$\overline{\mathrm{a}}$A h$_6 = $~$\overline{\mathrm{a}}$A$\overline{\mathrm{a}}$A/$\overline{\mathrm{a}}$A\underline{a}A, h$_7 =$~$\overline{\mathrm{a}}$A\underline{a}A/\underline{a}A$\overline{\mathrm{a}}$A, h$_8 =$~$\overline{\mathrm{a}}$A$\overline{\mathrm{a}}$A/\underline{a}A\underline{a}A, h$_9 =$~$\overline{\mathrm{a}}$AC\underline{c} (metastable), h$_{10} =$~$\overline{\mathrm{a}}$AC$\overline{\mathrm{c}}$ (metastable). The dotted horizontal line gives the average enthalpy for the stable HN configurations.}
\label{fig:stability}
\end{figure}

{\em The square network.} Fig.~\ref{fig:configs_diag}(a) shows the six possible configurations for a four-molecule unit cell that are non-equivalent by symmetry. Many examples for {\em all} of these six are found by the AIRSS. As mentioned previously, each configuration exists in distinct flat and rippled arrangements, with a first-order phase transition in $d$ between the two\cite{Zhao2014}. This is recovered naturally from the AIRSS, in agreement with previous studies\cite{Kaneko2013,Zhao2014,Kaneko2014,Zhao2014a}.

It is important to note that each molecule must donate one proton along a row of the grid and another along a column. Linear configurations (i.e., both protons along a single row or column) carry a significant energy penalty (they are found in some of the higher-energy defective cells). There is therefore both a decoupling of the ordering along rows and columns, and a long-range order imposed along a single row or column. As noted by Koga and Tanaka\cite{Koga2005}, the number of allowed configurations for $N$ molecules is $2^{2\sqrt{N}}$, which is not sufficient for a non-zero macroscopic entropy.

An important consequence of this is that, unless they are all energetically degenerate, all SN configurations correspond to separate phases, only one of them thermodynamically stable for any given set of external conditions. No transition from an ordered to a disordered phase on the same lattice is possible, unlike what happens in bulk between ice XI and ice Ih.

Fig.~\ref{fig:stability} shows the energetics of the different SN configurations. Surprisingly, the most stable is Ab/Cd, which is the only one with no net polarization (see Fig.~\ref{fig:configs_diag}(a)). This is confirmed for both the flat and rippled arrangements by the symmetry analysis of the relaxed crystal structures, discussed below. Conversely, the configuration with the largest polarization (Aa/Aa) is the highest in energy. The energetic ordering of configurations is constant for all $d$ and $P_l$. Therefore, our simulations predict the stable SN phases to be ordered and non-polar, and, hence, non-ferroelectric.

\begin{figure}[t!]
\includegraphics[width=0.49\textwidth]{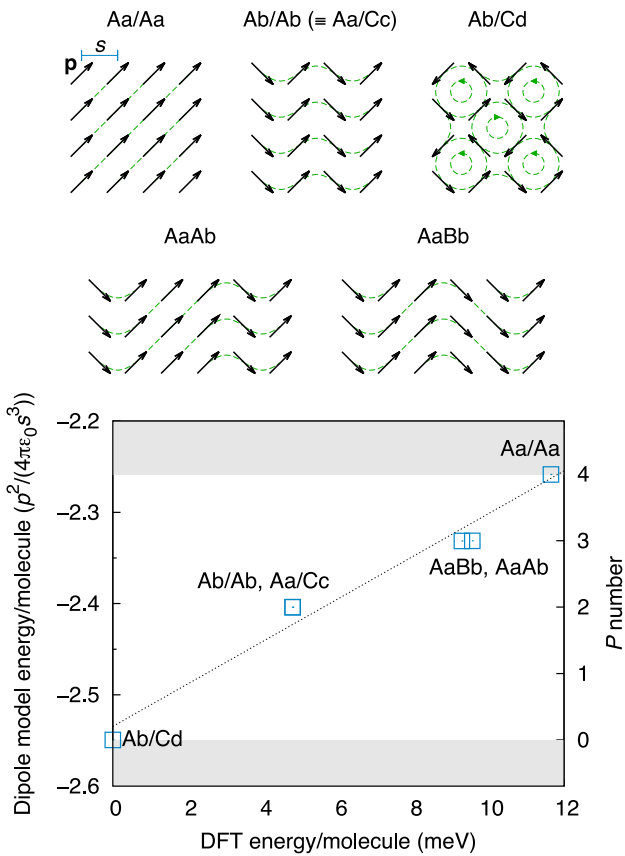}
\caption{{\bf Point dipole 2D lattice model for the four-molecule configurations of the square network.} The dashed lines give an idea of a continuous vector field defined by the dipoles. The electrostatic energy of the model is compared to the energies calculated by DFT at $d = 5$~\AA, $P_l = 0.01$~GPa$\cdot$\AA; the dotted line gives the best linear fit. The shaded areas enclose the allowed energy range for the model for any size unit cell $(0 \le P \le 4)$.}
\label{fig:dipoles}
\end{figure}

Despite going against previous predictions, the energetic ordering of configurations is readily explained by a simple electrostatic model of identical point dipoles on an ideal 2D periodic square grid (Fig.~\ref{fig:dipoles}). We have calculated the energy of the six configurations using this model, perfectly reproducing the DFT ordering and giving excellent estimates of relative energy differences, also shown in the figure.

In this model it is straightforward to see how every water molecule is represented as a dipole. However, it is also possible to approach the problem from a different limit, that of an idealized arrangement of symmetrized hydrogen bonds. In this case, dipoles can be obtained by considering the shift of each proton away from the bond centre, resulting in two smaller dipoles per water molecule. It can be shown that the two limits (protons close to O ions/protons close to bond centres) are perfectly equivalent; furthermore, a point charge model in which the protons are free to move between the two limits shows that the ratios in energy differences between configurations are constant at all points to a very good approximation (Supplementary Note~1). This goes some way towards explaining the robustness of such a simple model.

A close analysis of the dipole model (Supplementary Note~1) reveals that there is only a single feature accounting for practically all of the energy difference between configurations: the orientation of protons in neighbouring rows or columns of the SN grid (it is trivial to show by symmetry that energy differences are decoupled between rows and columns). The energetically favourable arrangement is for the two rows/columns to have opposite proton orientation; if instead it is the same, there is a relative energy penalty of:
\begin{equation}
E \left ( n \right )= \sum_{k=1}^\infty \left [ \frac{1}{\left ( k^2+n^2 \right )^{3/2}} - \frac{3 k^2}{\left ( k^2+n^2 \right )^{5/2}} \right ] + \frac{1}{2 n^3},
\end{equation}
where $n$ is the number of lattice spacings separating the two rows/columns on the lattice, and the energy scale is $p^2/\left ( 4 \pi \varepsilon_0 s^3 \right )$, determined by the magnitude of the molecular dipole $p$, and the lattice spacing $s$. For nearest neighbours ($n=1$), this gives a value of $7.28 \times 10^{-2}$~$p^2/\left ( 4 \pi \varepsilon_0 s^3 \right )$. For second-nearest neighbours ($n=2$), the value is already three orders of magnitude smaller; therefore, contributions from $n \ge 2$ can be safely neglected, and only nearest-neighbour interactions need to be considered. The schematic configuration diagrams from Fig.~\ref{fig:configs_diag}(a) can now be used to accurately estimate energy differences given by the dipole model (and, hence, by DFT), simply by counting the number of neighbouring rows and columns in the unit cell with the same orientation: four for Aa/Aa, three for AaAb and AaBb, two for Ab/Ab and Aa/Cc, and none for Ab/Cd.

More generally, the average number of parallel rows/columns per molecule ($P$ number) is a useful measure for classifying {\em all} allowed SN configurations independently of unit cell size. Its value can range between zero and four (i.e., the number of neighbours/molecule). Therefore, Ab/Cd must be the lowest-energy configuration for any size unit cell, and Aa/Aa the highest; it would be appropriate to rename them according to their $P$ number. A pair of parallel rows/columns can be seen as a defect of the $P = 0$ ground state, with a formation energy of $E \left ( 1 \right )$. Such a defect behaves similarly to a domain wall, which can propagate through the lattice but only be annihilated by another equivalent defect.

It is interesting to note that the $P = 0$ ground-state configuration has two unique features: (a) it provides the best possible screening of electrostatic interactions, leading to the fastest convergence of the dipole model energy with system size; and (b) it is the only one out of the six four-molecule configurations to exhibit topological charges, with the dipoles arranged into two square sublattices of vortices and antivortices (equal and opposite charge). All other configurations, instead, have infinite field lines for the coarse-grained vector field of dipoles. In general, sources and sinks are prohibited, as well as net electric charges.

We can therefore summarize the expanded ice rules governing the arrangement of water molecules in a SN monolayer as follows:
\begin{itemize}
\item The four-fold coordination of each molecule is still valid.
\item The presence of two short and two long OH distances for each O ion is also still valid, but there is an additional restriction forbidding linear molecular configurations (variants of this rule have already been noted by several authors\cite{Zangi2003a, Koga2005}). This restriction prevents proton disorder.
\item The energetic ordering of the allowed configurations depends linearly on the number of neighbouring rows/columns with the protons along both being oriented in the same direction.
\end{itemize}

\begin{figure}[t!]
\includegraphics[width=0.45\textwidth]{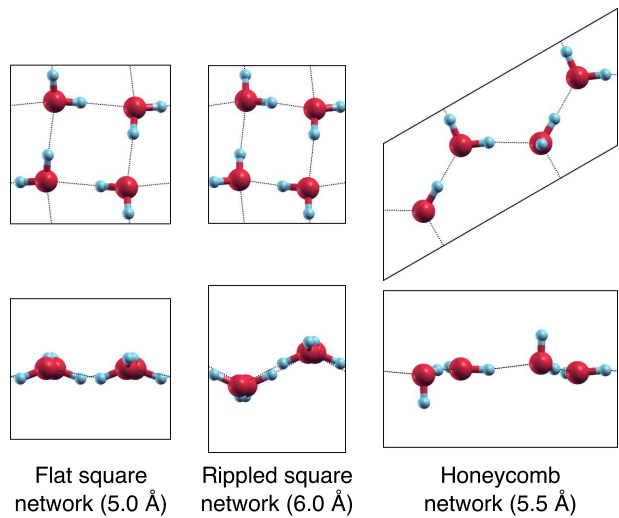}
\caption{{\bf Relaxed crystal structures.} The structures are shown in $x$--$y$ (above) and $x$--$z$ (below). The $d$ value is given in brackets; $P_l = 0.01$~GPa$\cdot$\AA\ for all cases.}
\label{fig:crystal}
\end{figure}

{\em The honeycomb network.} Graphene-like honeycomb monolayer ice cannot obey the bulk ice rules; the number of bonds/molecule ensures that half the molecules must donate two protons but receive only one, while the other half must donate one and receive two. This leads to $N/2$ dangling protons, which can be either above or below the monolayer plane (as shown in Fig.~\ref{fig:crystal}), and $N/2$ dangling lone pairs. Fig.~\ref{fig:configs_diag}(b) shows the ten non-equivalent configurations for a four-molecule unit cell (note that each of the five configurations shown has two possible placements of the dangling protons, both in the same direction or in opposite directions).

It is therefore reasonable to expect an energy penalty, but also a non-zero entropy, and, hence, a single disordered phase which is stabilized with temperature respect to the ordered SN phases. A naive estimation of the configurational entropy using Pauling's approach\cite{Pauling1935} gives $W = 2^{N/2} \times \left ( 9/2 \right )^{N/2} = 3^N$, where the first term accounts for the up/down position of the dangling protons, and the second for satisfying the ice rules for the in-plane bonds, and so $S = k_\mathrm{B} \ln{3}$. This is almost three times that of ice Ih. Given the enthalpy difference between the SN and HN configurations shown in Fig.~\ref{fig:stability}, this would be sufficient to obtain a transition between the two at room temperature.

\begin{figure}[t!]
\includegraphics[width=0.49\textwidth]{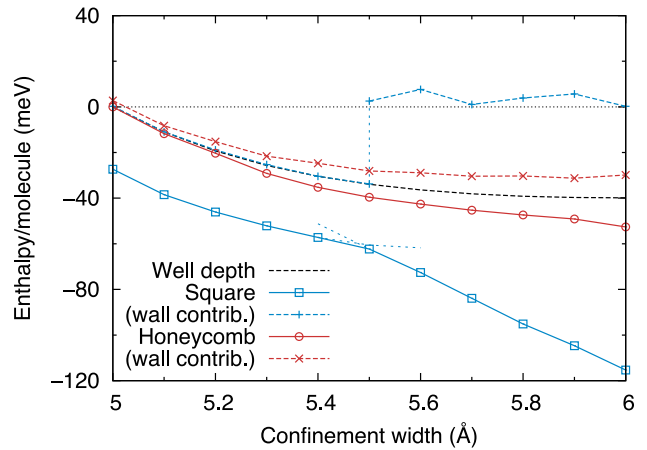}
\caption{{\bf Enthalpy as a function of confinement width.} The values shown are for $P_l = 0.01$~GPa$\cdot$\AA. For the SN there is a switch from the flat to the rippled phase, as shown by the dashed lines slightly extending the two segments into their metastable regions. The well depth gives the value of the confining potential at the centre of the confinement region. The contribution to the total enthalpy from this potential is given for each phase.}
\label{fig:enthalpy}
\end{figure}

\begin{figure*}[t!]
\includegraphics[width=0.98\textwidth]{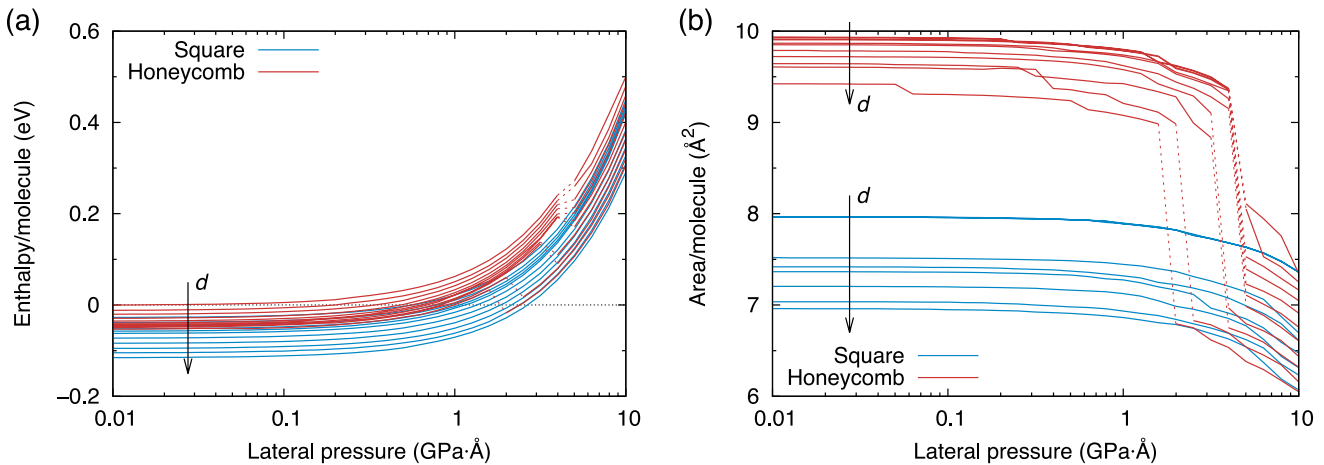}
\caption{{\bf Enthalpy and area as a function of lateral pressure.} This is plotted for varying widths of the confining potential, from 5 to $6$~\AA. Spontaneous phase transitions in the HN at high pressures are shown with a dashed line. The slight roughness in some of the area curves is due to the limitation in precision of the relaxation procedure, since the total energy is extremely insensitive to changes on this scale.}
\label{fig:area}
\end{figure*}

However, this is not the case. Indeed, contrary to the SN, the AIRSS finds only a subset of the allowed HN configurations. We test this systematically by building all ten unit cells by hand and relaxing them. The results confirm the fact that not all configurations are stable (details are given in Fig.~\ref{fig:configs_diag}(b)). The unstable ones either spontaneously relax to one of the stable ones, or remain trapped in a high-energy metastable state. Furthermore, the available data suggests that the instability arises from placing two in-plane or two out-of-plane molecules next to each other on the lattice; the stable configurations can then only be those for which the two types of molecules are segregated into the two triangular sublattices (i.e., the a and b sites). Interestingly, this results in neighbour pairs similar to the favourable configuration reported for polyhedral water clusters\cite{Kirov2008}. A further confirmation is obtained by testing two eight-molecule unit cells, one which obeys this restriction (aAaA/aAaA) and one which does not (abBA/BAab). The latter relaxes to the former, which is stable regardless of the placement of the dangling protons. We note that the energies of all the stable four- and eight-molecule configurations are almost degenerate, within 5~meV (Fig.~\ref{fig:stability}).

Interestingly, the stable configurations are {\em not} the ones expected by considering the possible arrangements of layers in bulk hexagonal ice. This is because in the bulk the dangling protons are locked into hydrogen bonds, while in the monolayer they are free to rotate. This gives rise to a mechanism by which two neighbouring molecules, one in and one out of plane, can both rotate in concert to swap the proton in the hydrogen bond connecting them. This is therefore a localized relaxation pathway for moving between configurations, which can be expected to have a small or even no energy barrier.

The reduction in the number of allowed configurations has a considerable impact on our previous estimate of the entropic contribution, which we can now revise to $W = 2^{N/2} \times \left ( 9/8 \right )^{N/2} = \left ( 3/2 \right)^N$, and so $S = k_\mathrm{B} \ln{3/2}$, equivalent to ice Ih. Consequently, the phase transition is shifted upwards to significantly higher temperatures ($\simeq$860~K), clearly no longer relevant for the solid phase.

Our conclusion, therefore, is that the HN phase is only ever metastable (we note that this analysis has been carried out at the most favourable point on the phase diagram for HN configurations). It is predicted to be disordered, and, hence, non-polar and non-ferroelectric. The lowest-enthalpy configuration, which should constitute an ordered phase at very low temperatures, cannot be identified with certainty; from Fig.~\ref{fig:stability}, it seems that relaxation effects in larger cell sizes might be crucial in stabilizing some configurations over others.

The ice rules governing the HN monolayer are starkly different both to the bulk and the SN:
\begin{itemize}
\item Each molecule must be three-fold coordinated, with half the molecules featuring out-of-plane dangling protons.
\item Two in-plane or two out-of-plane molecules cannot be neighbours in the network.
\item In-plane proton disorder is allowed, following the `two short/two long' rule for OH distances.
\item Proton disorder for the dangling protons (above/below the plane) is allowed with no restrictions, and is decoupled from the in-plane disorder.
\end{itemize}

\noindent {\bf Vibrational effects.} We have checked for any vibrational effects by calculating the phonon frequencies of the lowest-enthalpy four-molecule configurations for the two networks at $d = 5$~\AA, $P_l = 0.01$~GPa$\cdot$\AA\ (flat Ab/Cd for the SN, and $\overline{\mathrm{a}}$A\underline{a}A for the HN). We use the finite displacement method with a $\left ( 5 \times 5 \right)$ supercell and an ionic displacement of 0.02~\AA. The phonon DOS is shown in Supplementary Fig.~3.

The zero-point energy of the two monolayer configurations is almost identical, $111 \pm 1$~meV/molecule in both cases. This is $\sim$15\% smaller than the value reported for bulk ice at a similar level of theory\cite{Murray2012}. However, the vibrational free energy calculated from the phonon DOS decreases slightly faster for the SN as the temperature increases (e.g., a relative gain of 10~meV at $T = 430$~K). This further increases the stability of the SN phases at finite temperatures, thus reinforcing our previous conclusion on the metastability of the HN phase.

We have also checked that vibrational effects are very similar between SN configurations, and that switching from light to heavy water, despite lowering the zero-point energy by 26\%, is an entirely negligible effect when considering the relative stability of different configurations and phases.\\

\noindent {\bf Crystal structure.} We now examine in detail the ionic positioning within the unit cell, and the shape of the cell itself. We do so for a fine grid of points in a range of confinement widths from 5 to 6~\AA\ and lateral pressures from 0.01 to 10~GPa$\cdot$\AA.

Fig.~\ref{fig:enthalpy} shows how the enthalpy of the SN and HN phases depends on the confinement width for near-zero lateral pressure. The phase transition from the flat SN phase for small $d$ to the rippled SN phase for large $d$ can be observed at $\sim$5.5~\AA. The discontinuity is shown clearly by the contribution to the total enthalpy from the external wall potential; when viewed with respect to the well depth at the centre of the confinement region (also plotted), this gives a measure of the amplitude of the monolayer in $z$. While the HN follows this baseline value quite closely, thereby remaining fairly flat for all $d$, the SN switches discontinuously between an almost perfectly flat arrangement and a rippled one. The ripples are seen in the $z$-position of the O ions for alternate rows of molecules, as shown in Figs.~\ref{fig:configs_diag}(a) and \ref{fig:crystal}. There is a clear gain in internal energy from buckling (related to the improved alignment of the hydrogen bonds), which competes with the penalty from the confining potential; hence, the transition when increasing $d$.

In terms of Bravais lattices, the flat SN phase is indeed perfectly square within the relaxation tolerance, as should be expected from the symmetry of the bonding arrangement; the layer group is {\em p$\overline{4}$b$2$} ({\em p$4$/mbm} for the O sublattice). Once ripples are introduced, the symmetry is reduced; the rippled SN phase is therefore slightly rectangular in practice, with an $a/b$ ratio of 0.9 at $d = 5.5$~\AA, $P_l = 0.01$~GPa$\cdot$\AA, and decreasing with both increased $d$ and $P_l$; the layer group is now {\em p$2_1$/b$11$} (equivalent for the O sublattice). Nearly all configurations from Fig.~\ref{fig:configs_diag}(a) follow a similar pattern; the only two exceptions, Aa/Aa and AaAb, are the ones in which there is a non-zero net polarization which is not parallel to one of the bonding directions. In these cases, the unit cell switches from rhombic (flat) to oblique (rippled). The distortion in the unit cell angle with respect to the square/rectangular configurations is on the order of 1--10\%. For Aa/Aa, the layer group of the flat arrangement is {\em p$2$11} ({\em cmmm} for the O sublattice), and that of the rippled arrangement is {\em p$11$a} ({\em p$112$/a} for the O sublattice).

The HN phase is close to hexagonal in symmetry for all $d$ and $P_l$, and for all stable configurations from Fig.~\ref{fig:configs_diag}(b). Distortions are small, on the order of 1\% in the unit cell angle, and 1--10\% in the $a/b$ ratio. Since the phase is disordered, only the symmetry of the O sublattice averaged over all configurations is relevant; the layer group is therefore {\em p6/mmm}.

Fig.~\ref{fig:area} shows how the enthalpy and area/molecule vary with the confinement width and lateral pressure for the SN and HN phases (note that the area of the flat SN phase is almost completely insensitive to $d$, and that there is a discontinuous transition to the rippled SN phase, which can be seen in the large gap between curves for the SN in Fig.~\ref{fig:area}(b)). An interesting phenomenon is observed: the metastable HN phase becomes unstable above a critical pressure, and spontaneously relaxes to a SN configuration. The critical pressure decreases with $d$. This explains why the AIRSS only finds HN configurations for small values of $d$ and $P_l$.

\begin{figure}[t!]
\includegraphics[width=0.49\textwidth]{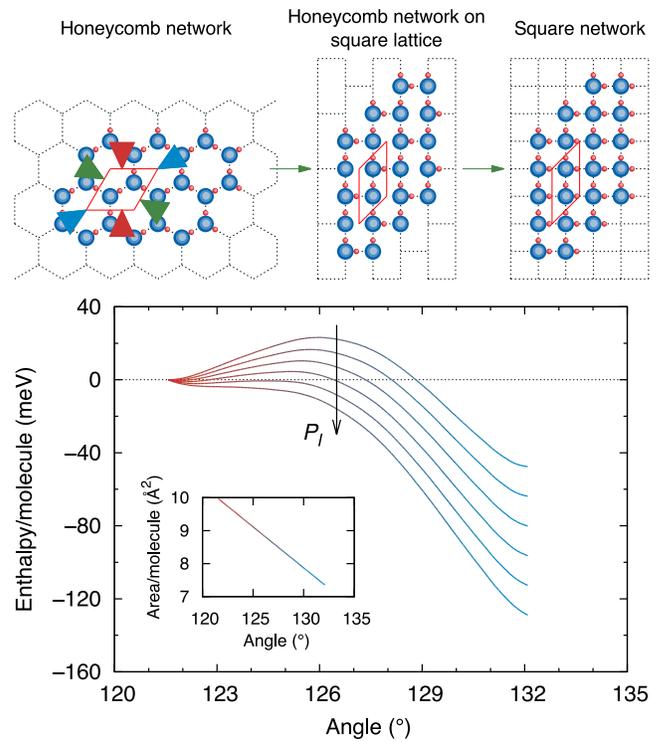}
\caption{{\bf Instability of the honeycomb network at high isotropic lateral pressures, leading to a spontaneous relaxation to a square network configuration.} The three possible strains to the hexagonal cell are shown by the pairs of green/red/blue arrows; the path given by the green one is followed here. The simulations are performed at $d = 6$~\AA. The pressure is varied linearly from 0 to 5~GPa$\cdot$\AA, and all the curves are aligned to zero for the unstrained cell.}
\label{fig:hex2square_diag}
\end{figure}

There are three possible strains which transform a hexagonal cell to a square one (Fig.~\ref{fig:hex2square_diag}). Using the cell angle as the order parameter, we follow this transformation for a representative confinement width, and show the effect on the enthalpy caused by varying the lateral pressure. The behaviour is that of a first-order phase transition; therefore, what is observed is the crossing of the phase stability limit. This also suggests the existence of a phase transition from the SN to the HN, although the critical pressure is in the negative regime.

It is also interesting to note that the distortion can be seen to have an intermediate stage, that of molecules arranged on a square or quasi-square lattice but maintaining the honeycomb bonding network, as illustrated in Fig.~\ref{fig:hex2square_diag}. This is because half of the molecules in the HN configuration need to rotate to bring their dangling proton in plane and complete the SN bonding. This intermediate configuration can itself sometimes be metastable (especially for small values of $d$), and, indeed, accounts for some of the previously mentioned low-lying defective SN configurations found by the AIRSS. It is therefore reasonable to assume that this mechanism will be important for introducing a degree of disorder in the SN phases at moderately high temperatures, and as a way of propagating changes between different configurations through the lattice.\\

\noindent {\bf Electronic structure.} In order to examine how the electronic charge is distributed around the molecules, and how this differs with respect to bulk water, we turn to a maximally-localized Wannier function\cite{mlwf} (MLWF) analysis of the system. The MLWFs are obtained from our DFT calculations by interfacing\cite{Korytar2010} with the Wannier90\cite{Mostofi2008} post-processor code.

\begin{figure}[t!]
\includegraphics[width=0.49\textwidth]{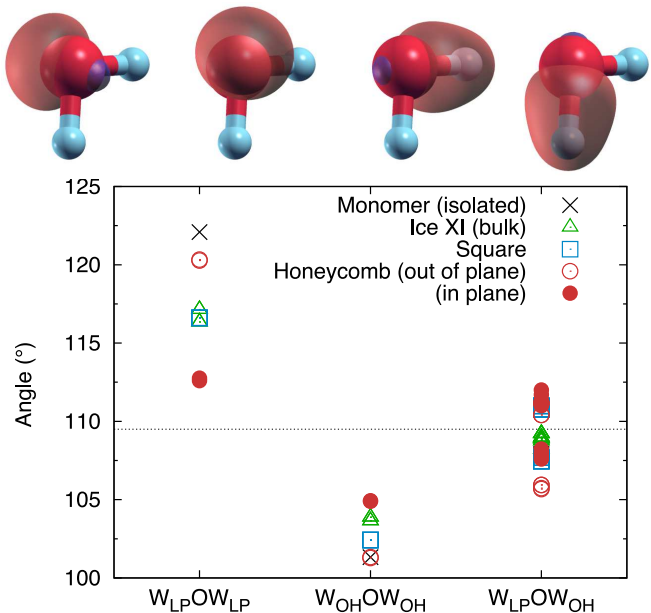}
\caption{{\bf Angle between two MLWF centres and the O ion.} The MLWFs are divided into lone pairs (LP) and OH bonds (OH). The dotted line shows the ideal tetrahedral angle. Contour plots for the four MLWFs of an individual molecule are shown above; the positive/negative components of the contour are shown in red/blue.}
\label{fig:mlwf}
\end{figure}

Fig.~\ref{fig:mlwf} shows the typical MLWFs for the water molecule. The molecular bonding is very strong, so we cannot expect a qualitative change in this picture; however, subtle changes in the electronic structure will be reflected in the position of the MLWF centres and their spreads. We analyze the monolayer SN and HN phases, as well as bulk ice XI and an isolated water molecule. We note that the monolayer results are very similar for the different configurations of Fig.~\ref{fig:configs_diag}, and also for the flat and rippled SN arrangements.

In several respects the monolayer MLWFs are found not to differ significantly from the bulk crystal. This is the case for the spreads: $0.49 \pm 0.01$~\AA$^2$ for the OH bonding orbitals, and $0.58 \pm 0.01$~\AA$^2$ for the lone pairs. (The values are somewhat lower for the isolated molecule: 0.47 and 0.52~\AA$^2$, respectively.) The distances between the WF centres and the O ion are also very similar, including for the isolated molecule: $0.50 \pm 0.02$~\AA\ for the OH bonding orbitals, and $0.32 \pm 0.02$~\AA\ for the lone pairs.

The main difference is found for the HN monolayer when considering the angle between pairs of MLWFs and the O ion. This is shown in Fig.~\ref{fig:mlwf}. Interestingly, the values obtained for the SN phases are very similar to those of bulk ice. Within the level of detail offered by the MLWFs, therefore, the electronic structure of the water molecules can be considered not to be altered significantly between these phases. The HN monolayer, however, is more complicated. First of all, there is a noticeable difference between the out-of-plane and in-plane molecules. The MLWFs of the former are positioned similarly to those of an isolated molecule, as can be seen in the figure (this is true for all of the molecule's MLWFs, not only the one associated with the dangling proton). The MLWFs of the latter are positioned differently from all other phases. In particular, the angle between the two lone-pair orbitals is reduced; this can be understood to favour the formation of a hydrogen bond pointing between the two, as is required for the quasi-planar sp$^2$-like bonding of the honeycomb lattice.

The magnitude of the molecular dipole moment is found to be quite insensitive to this angle: it is $3.0 \pm 0.04$~D for {\em all} molecules in the SN and HN phases, compared to $3.4 \pm 0.06$~D for ice XI and $2.0 \pm 0.06$~D for the isolated molecule (additional calculations show that there is a consistent overestimation of 5--10\% due to the limitations of the basis set). It is also important to note that the changes in the angle between lone-pair orbitals are not correlated with the HOH angle. Therefore, we expect the HN monolayer to pose a particularly interesting challenge for force-field modelling.

\section*{Discussion}

The picture offered by first principles simulation differs significantly from the previous review of empirical force-field results\cite{Zhao2014a}, mainly based on data obtained with the TIP5P model. The overall phase diagram is greatly simplified by ruling out the existence of any stable phase other than the SN ones. We note that the previously reported\cite{Bai2010,Zhao2014} Archimedean tiling phase could not be recovered from our AIRSS due to the limitation in unit cell size; therefore, we have checked this separately by relaxing the ionic configuration reported by Zhao {\em et al.}\cite{Zhao2014}, finding it to be high in enthalpy (44~meV/molecule above the ground state in Fig.~\ref{fig:stability}, at the same level of the metastable HN configurations), thus not affecting our conclusions. As with the HN configurations, the double hydrogen bond arrangement between pairs of atoms predicted by TIP5P is found to be unstable, with our relaxation resulting in one dangling proton per two molecules.

The SN and HN phases are also found to differ from these previous results in terms of their proton ordering, crystal symmetry and net polarization. It is particularly noticeable that our prediction for the stable flat SN phase is in perfect agreement with recent experimental observations\cite{Algara-Siller2015}, both in terms of the square symmetry of the O sublattice and the value of the lattice parameter (2.82~\AA, compared with $2.83 \pm 0.03$~\AA\ from experiment). The first principles approach, therefore, successfully untangles the contradictory predictions from different force-fields models; furthermore, the point dipole 2D lattice model provides a simple physical explanation as to why the Ab/Cd configuration is the most stable.

It is surprising that previous force-field studies do not all recover the correct ordering of SN configurations (with some models placing the most polar and rhombic one, Aa/Aa, as the most stable instead of the least stable). This is unexpected because the striking agreement between our simple dipole model and DFT suggests that the ordering is entirely dominated by electrostatics, which all force-field models should have no problem in reproducing. Indeed, we have confirmed this to be the case by comparing the electrostatic interaction of different configurations of water dimers placed on the SN for DFT and a variety of force-field models (TIP3P, TIP4P, TIP5P). The value for DFT is approximated by a seven-site point charge model, using the three ionic positions and the four MLWF centres. Both the DFT and force-field models are in excellent agreement with the dipole model.

However, previous studies using TIP5P show that the relaxation of different configurations is not consistent with DFT. In particular, the polar Aa/Aa configuration is severely distorted from the idealized square lattice, with a unit cell angle of $\sim$70\%. Such a distortion significantly lowers the electrostatic energy of Aa/Aa with respect to Ab/Cd, reversing the ordering of configurations. For DFT, instead, the rhombic distortion is much smaller, and so the dipole model remains accurate.

In general, our results suggest that TIP5P suffers from an over-emphasis on the tetrahedral bonding, resulting in overly distorted SN configurations, as well as the incorrect bonding for HN configurations. Results for other force-field models are much more limited, although there is some evidence that three- and four-site models correctly order the SN configurations\cite{Zhao2014, Zhao2014a, Algara-Siller2015}, which is consistent with their reduced emphasis on the tetrahedral bonding geometry.

Another important point to make is that the results presented here, in particular the relative stability of the SN and HN phases, will depend on the detailed form of the confining potential. We have purposefully chosen one of the simplest possible confinement schemes in order to investigate the intrinsic tendencies of monolayer ice, without bias from a particular chemical environment. Therefore, our results offer a comprehensive characterization of the most relevant possibilities for water to crystallize in a monolayer, which will then be subjected to different such biases in different realistic scenarios.

In particular, several reasonable modifications to the confining potential might favour the HN over the SN, e.g., adding an attractive term to the H ions, and/or introducing a periodic modulation to the walls with a quasi-commensurate hexagonal lattice, typical of many surfaces. Such modifications can be expected to decrease the energy penalty of the HN, and shift the first-order phase transition with the SN towards positive pressures. Indeed, various specific cases have been shown to favour a HN configuration\cite{Feibelman2004, Carrasco2012, Li2012, Mosaddeghi2012, Ferguson2012}. It is however interesting to note that the only available experimental results show the intrinsic tendency towards the SN phase to overcome the hexagonal confining potential provided by graphene sheets\cite{Algara-Siller2015}.

The formation of monolayer ice requires very small, sub-nm confinement regions; it is therefore reasonable to question the possibility for experimental synthesis, or for natural formation in biological or geological structures. A simple thermodynamic argument, combined with predictions from DFT, suggests that only moderately high pressures are necessary to cause the insertion of ice at equilibrium between sheets with such narrow spacing. We reach this conclusion by comparing the chemical potential of water molecules in bulk ice XI under isotropic pressure and our most stable nanoconfined phases; the two are equilibrated at a value of the external pressure of $\sim$1~GPa for $d = 6$~\AA\ (increasing to $\sim$2~GPa for $d = 5$~\AA). The value will depend both on the confinement width and the form of the confining potential, which defines the well depth (as shown in Fig.~\ref{fig:enthalpy}) with respect to the zero of potential outside the confinement region. A strongly hydrophilic or hydrophobic surface will shift the well depth, and, hence, the external pressure required for insertion.

\footnotesize{\section*{Methods}

\noindent {\bf Simulation setup.} All the simulations are performed with the SIESTA\cite{Soler2002} DFT code, with norm-conserving pseudopotentials in Troullier-Martins form\cite{Troullier1991} and a basis set of finite-range numerical atomic orbitals\cite{Junquera2001} (NAOs). The pseudopotentials are the same as those described in a previous study of liquid water and ice\cite{Corsetti2013}. We employ a variationally-obtained\cite{Anglada2002} double-$\zeta$ polarized NAO basis, also used in previous studies\cite{Wang2011, Corsetti2013, Corsetti2013b}. This basis has been extensively tested against both larger NAO bases and plane-wave calculations\cite{Corsetti2013, Corsetti2013b} (in these studies, it is referred to as \DZPmv); it has shown a high degree of transferability, and accuracies comparable to plane-wave cutoffs of $\sim$850~eV for energy differences and ionic forces, and $\sim$1500~eV for absolute pressures. SIESTA represents the electronic density on a real-space grid, for which we use a cutoff of 2040~eV (150~Ry).

For xc, we use the vdW-DF functional by Dion {\em et al.}\cite{Dion2004}, but substituting the revPBE exchange energy with PBE, as previous studies have shown this to be among the best possible descriptions currently available for water from first principles\cite{Wang2011, Zhang2011a, Corsetti2013b}.

The confining walls are described by a classical Lennard-Jones (L-J) 9-3 potential with parameters $\sigma = 2.5$~\AA\ and $\varepsilon = 13.0$~meV (1.25~kJ/mol), applied only to the O ions; the separation between the origin of the two walls along $z$ is referred to as $d$. This potential is structureless in the $x$--$y$ direction, and was originally developed to approximate a hydrophobic paraffin-like surface\cite{Lee1984}. We chose it due to its simplicity, and because it is the same one used in many previous studies, thus allowing for a direct comparison of results\cite{Kumar2005, Koga2005, Bai2010, Kaneko2013, Zhao2014, Kaneko2014, Zhao2014a}. Tests on our system show that the results are not significantly affected by the details of the confining potential, in agreement with the study of Kumar {\em et al.}\cite{Kumar2007}.

The simulation cells are periodic in all directions; we add a buffer region of 11~\AA\ in $z$, and use a dipole correction for slab geometries to prevent artifacts. We use a Monkhorst-Pack (MP) k-point sampling grid\cite{mp_grid} in the $x$--$y$ plane only, with a length cutoff of 10.6~\AA\ (20~a$_0$) as defined by the scheme of Moreno and Soler\cite{Moreno1992}. Convergence tests were performed for the k-point sampling, size of the buffer region, and real-space cutoff, to ensure an accuracy within 1~meV/water molecule.

\noindent {\bf Structural relaxations.} Relaxations are performed at particular values of the 2D lateral pressure (defined below), with no restrictions on either the ionic positions or the two vectors defining the 2D unit cell (the third vector in $z$ is held constant and perpendicular to these). The convergence tolerances used are 1~meV/\AA\ for the maximum ionic force, and 0.1~MPa for the maximum error on a component of the stress tensor. A robust and reliable relaxation procedure is achieved by cycling repeatedly through 20 steps of conjugate-gradient minimization, 20 steps of a modified Broyden optimization\cite{Johnson1988}, and 20 steps of the FIRE algorithm\cite{Bitzek2006}, until reaching convergence.\\

\noindent {\bf Lateral pressure definition.} The 2D lateral pressure $P_l$ which we refer to throughout the paper is given in dimensions of [pressure]$\cdot$[length] (GPa$\cdot$\AA). This is because, for the ground-state calculations performed here, the conventional components of the stress tensor are ill-defined due to the arbitrariness in setting a length for the $z$ dimension of the unit cell. An approach based on the virial\cite{Kumar2005} is also not feasible, as we do not perform dynamic simulations. Therefore, we use a $\left ( 2 \times 2 \right )$ 2D stress tensor with components $\sigma^{2\mathrm{D}}_{ij} = l_z \sigma^{3\mathrm{D}}_{ij}$, where $\left \{ i,j \right \} = \left \{ x,y \right \}$ and $l_z$ is the length of the unit cell vector in $z$ used for the simulation. The 2D lateral pressure is then $P_l = -\partial E/\partial A$, where $E$ is the ground-state energy of the system, and $A$ is the area of the unit cell in $x$--$y$. A simple estimate of the 3D lateral pressure can be obtained by using the confinement width as a reasonable approximate $z$-width for the system: $P_l^{3\mathrm{D}} \approx P_l^{2\mathrm{D}}/d$.

\section*{Acknowledgements}

\noindent This work was partly funded by grants FIS2012-37549-C05 from the Spanish Ministry of Science, and Exp.\ 97/14 (Wet Nanoscopy) from the Programa Red Guipuzcoana de Ciencia, Tecnolog\'{i}a e Innovaci\'{o}n, Diputaci\'{o}n Foral de Gipuzkoa. We thank Richard Koryt\'{a}r and Javier Junquera for their work on the SIESTA interface to Wannier90, and Raffaele Resta and M.-V. Fern\'{a}ndez-Serra for useful discussions. The calculations were performed on the arina HPC cluster (Universidad del Pa\'{i}s Vasco/Euskal Herriko Unibertsitatea, Spain). SGIker (UPV/EHU, MICINN, GV/EJ, ERDF and ESF) support is gratefully acknowledged.

\section*{Author contributions}

\noindent F.C.\ designed the initial research project. P.M.\ and F.C.\ performed the AIRSS simulations; F.C.\ performed all other simulations. F.C.\ and E.A.\ analysed the data and interpreted the results. F.C.\ wrote the manuscript; all authors edited the manuscript.

\section*{Additional information}

\noindent {\bf Competing financial interests:} The authors declare no competing financial interests.

\section*{References}}

\end{document}